\input{aipcheck}

\documentclass[
    ,final            
  ]
  {aipproc}

\newcommand{\Sleuth}{{\sc Sleuth}}
\newcommand{\sumPt}{{\ensuremath{\sum{p_T}}}}
\newcommand{\scriptP}{\ensuremath{{\cal P}}}

\def \met {{\ensuremath{\,/\!\!\!\!E_{T}}}}

\def \lumi {2.0~fb$^{-1}$}

\layoutstyle{8x11double}

\begin{document}

\title{Global Search for New Physics at CDF}


\author{Si Xie on behalf of the CDF collaboration}{
  address={Massachusetts Institute of Technology,
Cambridge, MA 02139, USA.}
}

\begin{abstract}
A model-independent global search for new physics has been performed at the CDF experiment. This search examines 399 final states, looking for discrepancies between observation and the standard model expectation in populations, kinematic shapes, bumps in mass distributions suggestive of new resonances and the tails of the summed transverse momentum distribution. This global search reveals no evidence of physics beyond the Standard Model in \lumi\ of $p\bar{p}$ collisions at $\sqrt{s}=1.96$~TeV.

\end{abstract}

\maketitle


\section{Introduction and Strategy}

\begin{figure}
$\begin{array}{c}
\includegraphics[width=0.45\textwidth,angle=0]{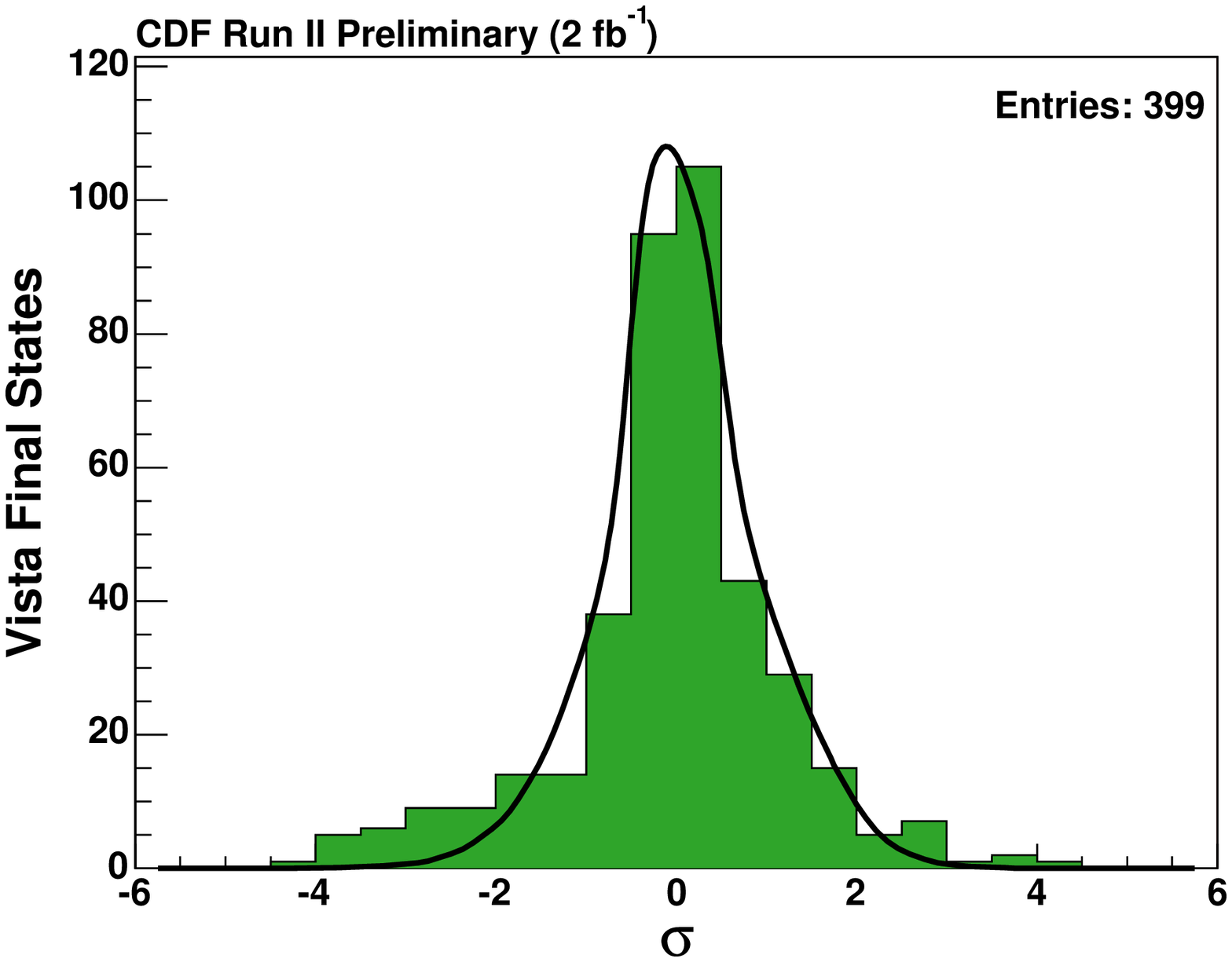} \\
(a) \\
\includegraphics[width=0.45\textwidth,angle=0]{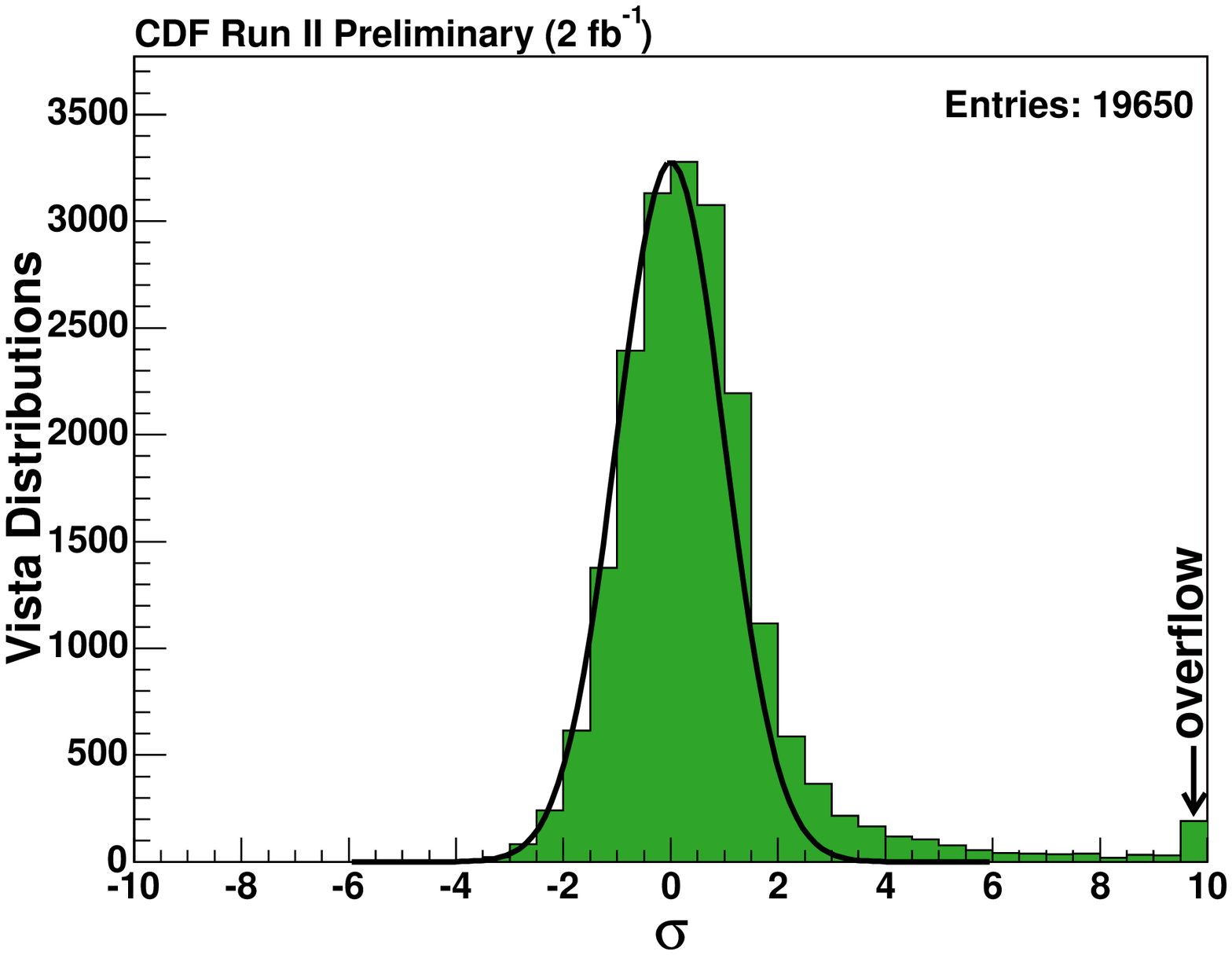} \\
(b)
\end{array}$
\caption{Summary of the results of the global comparison of data to SM prediction for populations of final states (a) and shapes of kinematic distributions (b). The black line represents the theoretical expectation assuming no beyond-SM physics.}
\label{fig:vistaSummary}
\end{figure}

\begin{figure}
  \includegraphics[width=0.45\textwidth,angle=0]{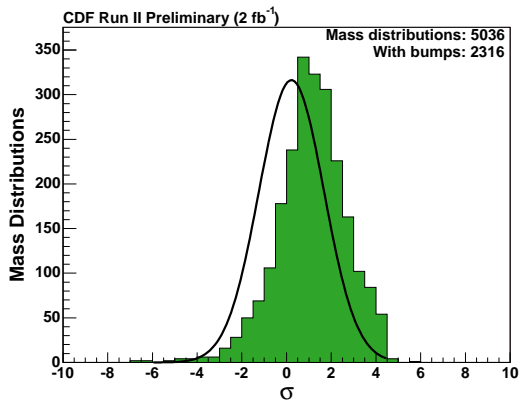}
  \caption{Significance of the most interesting bump in each mass variable.  Each entry corresponds to one mass distribution found to contain at least one bump satisfying the quality criteria.}
  \label{fig:vistaSummary_bump}
\end{figure}

\begin{figure}

$\begin{array}{c}
\includegraphics[angle=-90,width=0.40\textwidth]{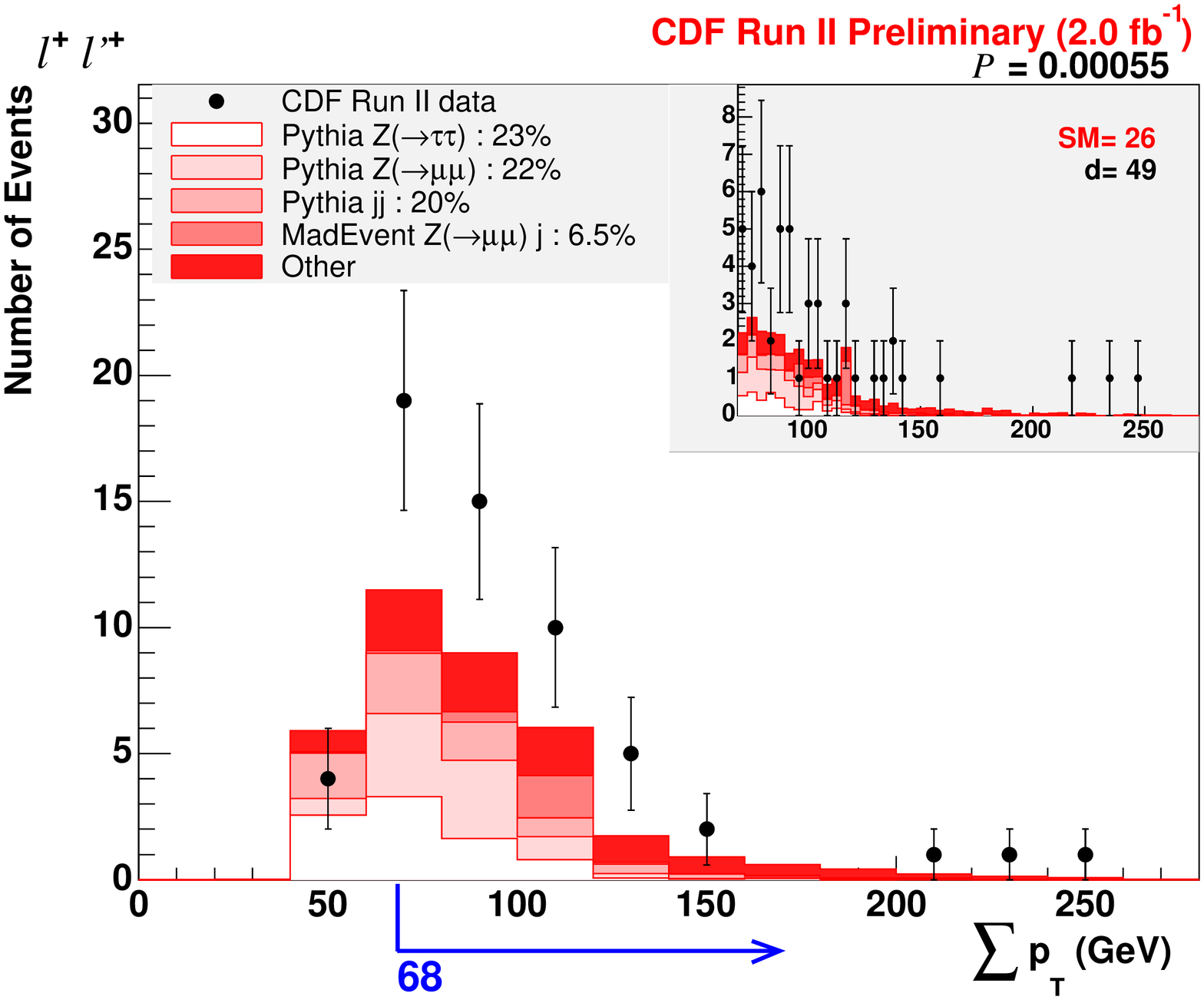} \\
(a) \\
    \includegraphics[angle=-90,width=0.40\textwidth]{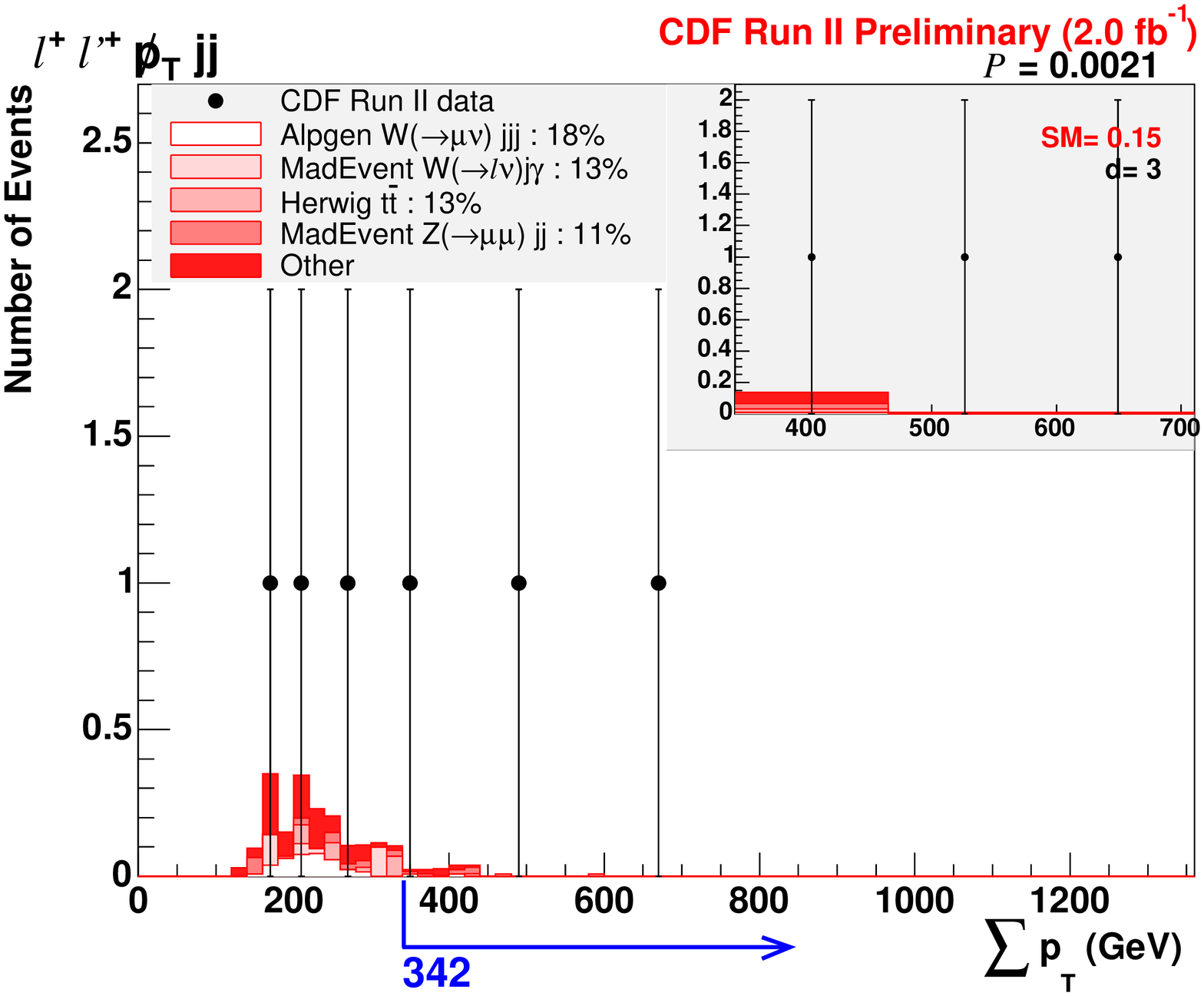} \\
(b) \\
    \includegraphics[angle=-90,width=0.40\textwidth]{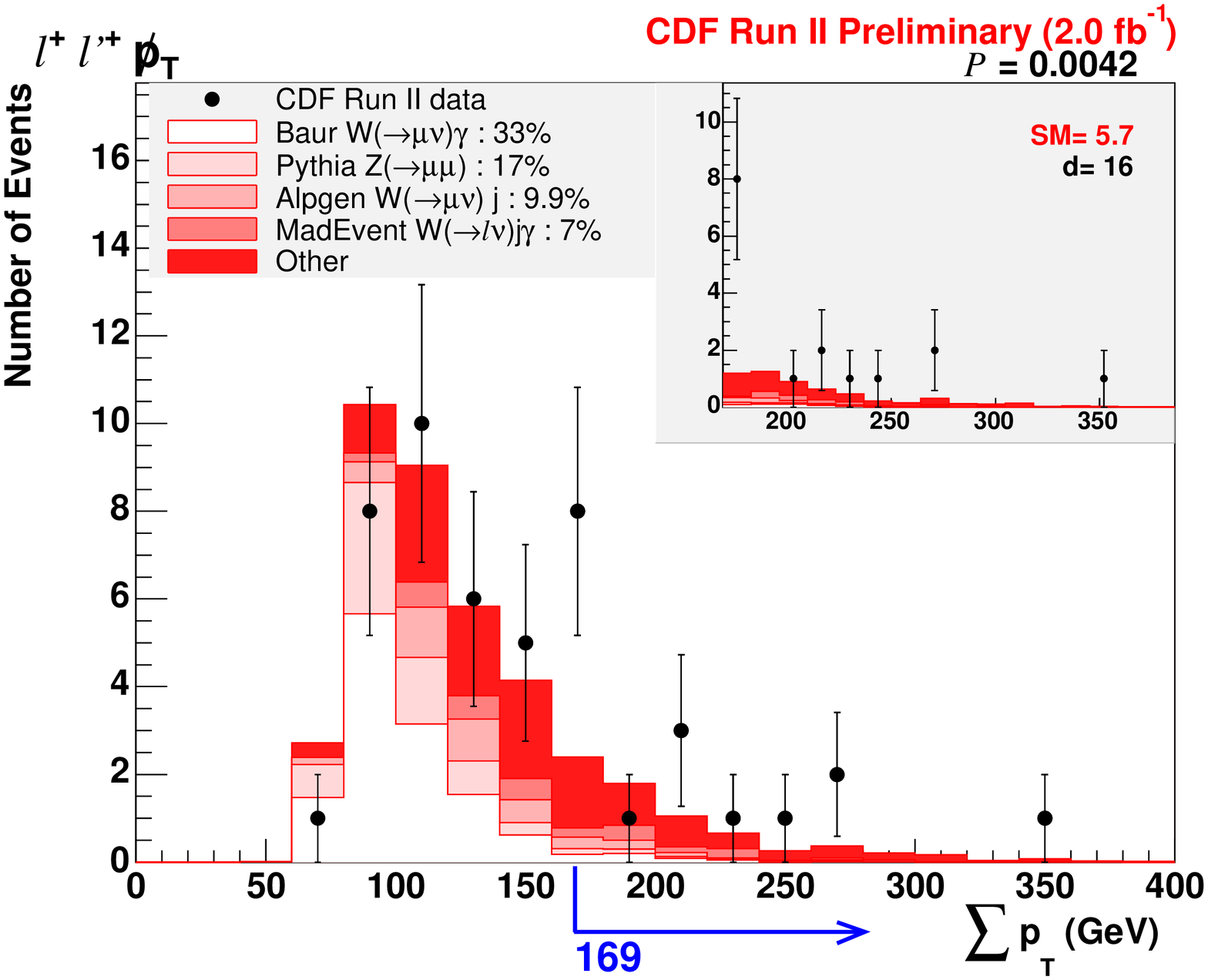} \\     
(c) 
\end{array}$

  \caption{The three most interesting final states found by \Sleuth\ in \lumi.
The label in the top left corner of each plot shows the identified objects present in each final state, where {\em l}$^\pm$ is a lepton ($e$ or $\mu$), {\em l'} is an additional lepton of different flavor, $j$ denotes a jet and ${\not\!\!p_T}$ represents missing transverse momentum. The region identified as having the most significant excess of data over background is indicated by the blue line and displayed in the inset. The significance of the excess is shown by the value of \scriptP\ in the top right corner.}
  \label{fig:sleuthPlots}
\end{figure}

A model-independent global search for new physics has the unique ability to complement conventional dedicated searches for physics beyond the Standard Model. Due to the current proliferation of extensions to the Standard Model, it is extremely difficult to decide among the large number of candidate models which one is the most favored. Even within the Minimal Supersymmetric Standard Model, which is arguably the most favored model in the particle physics field, there still remains the problem of covering the entirety of the 107 dimensional parameter space in any search meant to decide for certain whether supersymmetry is a true symmetry of nature. Thus, this global, model-independent search attempts to cover the large remaining regions of signature space, to ensure that no possible hint of new physics is missed.

The overall strategy of this global search is to maximize the scope of data that is analyzed, use our expertise of the Standard Model to implement a global Standard Model prediction, and then to focus attention and scrutiny on genuine statistically significant outlying effects. To achieve these goals, a predominantly Monte Carlo based Standard Model prediction is generated. Rather than adopting predefined signal and control regions, the entirety of the data are assumed to potentially contain the first signs of new physics. At the same time, the entirety of the data contribute information for constraining the small number of physically well motivated correction factors, which modify the Standard Model prediction. These adjustments are designed to give an accurate accounting of the degree to which we understand systematic effects. Those discrepancies which persist after these adjustments are considered as possible candidates for evidence of new physics, and are subsequently further investigated and scrutinized. Those candidates for which one has reasonably ruled out an explanation in terms of a statistical fluctuation, a detector effect, or background mismodelling, will provide the basis of claims of new physics. Further details of this global search may be obtained in a previous publication based on 927 pb$^{-1}$ of data~\cite{vista:prd}.

\section{Vista}

This global analysis begins by identifying the objects of interest. Electrons, muons, photons, taus, jets, jets tagged as originating from a bottom quark, and missing transverse momentum \met\ with sufficiently large transverse momentum ($p_T > 17$ GeV) are identified and reconstructed using standard CDF criteria. A set of global event selection criteria is imposed in order to retain a wide spectrum of events at the highest transverse momenta for analysis. The events are partitioned into exclusive final states based on the objects that they contain. 
A large set of observables relevant to each final state is constructed from kinematic variables. They include object transverse momenta, polar and azimuthal angles, angles between pairs of objects, masses of all combinations of objects, and a number of more specialized observables. 

To construct the Standard Model prediction, a diverse set of Monte Carlo generators are used to generate events to cover the entire spectrum of particle production at the Tevatron. They include dijet production, photon+jets production, diphoton production, production of a vector boson in association with jets, production of a pair of vector bosons, and top quark pair production. Extremely careful accounting of the multitude of backgrounds must be made in order to ensure that no backgrounds are missed and none are double counted. These background events are subsequently reweighted according to a set of correction factors fit from data to take into account systematic effects. They include the integrated luminosity, the k-factors of various Standard Model processes, trigger and reconstruction efficiencies, and a number of fake rates~\cite{vista:prd}. The values of these factors are obtained by maximizing the global agreement with data in the populations of the exclusive final states, subject to external constraints from theory calculations and other experimental results. The degree to which any particular discrepancy may be resolved by an adjustment of the correction factors simply reflects the degree to which systematic effects are not well understood enough to constrain those factors.

Having implemented the entire Standard Model prediction, it is possible to make a global comparison between the data and the prediction. Using Poisson statistics and the Kolmogorov-Smirnov test, the Vista global comparison quantifies the degree to which the populations of the exclusive final states and the shapes of the kinematic observables agree with the prediction. The results are shown in Figure \ref{fig:vistaSummary}. After accounting for the trials factor associated with searching in multiple final states, we find no statistically significant discrepancy from the Standard Model prediction in the populations. Of the 19650 kinematic observables that are investigated, 555 actually do exhibit a statistically significant discrepancy. However, further investigation and analysis of those effects reveal that they likely arise from a difficulty in modelling soft QCD jet emission in the underlying Monte Carlo events. In particular, they are attributed to the fact that the Pythia showering parameters are not particularly well tuned to the Tevatron data~\cite{vista:prd}. 

\section{Bump Hunter}
Having gained confidence in the Standard Model background prediction from the results of the Vista global comparison, we are able to use more specialized statistical tools to look for new physics. An algorithm called the Bump Hunter searches in a total of 5036 mass variables to find local excesses of data over background suggestive of resonant production of new particles~\cite{georgios_thesis}. All mass distributions are scanned in windows whose sizes are consistent with the detector resolution, implicitly assuming that the intrinsic width of the new particle is narrower by comparison, and the window with the most significant excess of data over background is selected. Certain quality criteria are imposed to eliminate pathological cases. The results are summarized in Figure \ref{fig:vistaSummary_bump}. The shift between the expectation from pseudo-experiments and the observed data is due to the fact that this algorithm is sensitive to local mismodelling in the background, illustrating the well known fact that the Monte Carlo events do not perfectly reproduce the data. There is one genuinely outlying excess, which warranted further scrutiny. This bump has a significance of $4.1 \sigma$ after accounting for the trials factor, and occurs in the mass of all jets in the 4-jet final state. The region that is selected coincides with the region in which the well established soft jet emission discrepancy lies. As a result, one concludes that the effect is likely due to the same underlying QCD parton showering issue and is unlikely to be new physics.

\section{Sleuth}

Finally, an algorithm called \Sleuth\ attempts to find statistically significant excesses in the tail of the \sumPt\ distribution. Each semi-infinite region beginning with each data point and extending to infinity are considered and the one exhibiting the most significant excess of data over background is selected in each final state. 

The three most significant \Sleuth\ final states are shown in Figure \ref{fig:sleuthPlots}. The significance of the top \Sleuth\ discrepancy is 0.00055. After accounting for the trials factor, this significance translates to a final probability of 8\% that hypothetically similar CDF experiments would yield a final state that is as or more interesting than the most interesting final state found. 

\section{Conclusions}
CDF has completed a model-independent global search for new physics in \lumi\ of data. The populations of the 399 final states exhibit no statistically significant discrepancies relative to our Standard Model implementation. Of the 19650 kinematic observables that are examined, 555 show clearly significant deviations from the prediction. Most of these are found to be due to difficulties in modelling soft QCD jet emission and in tuning the Pythia showering parameters to the Tevatron data, which also explains the only significant outlier in the search for mass resonances in 5036 mass distributions. The small number of remaining shape discrepancies is understood to be due to the overall crudeness of the correction model. Finally, the \Sleuth\ statistic, while pointing out the interesting signature of same sign, opposite flavour dilepton events in the top three final states, revealed no statistically significant excess at high $\sumPt$. Unfortunately, this analysis is unable to make any discovery claims of new physics.




\bibliographystyle{aipproc}   

\bibliography{sample}

\IfFileExists{\jobname.bbl}{}
 {\typeout{}
  \typeout{******************************************}
  \typeout{** Please run "bibtex \jobname" to optain}
  \typeout{** the bibliography and then re-run LaTeX}
  \typeout{** twice to fix the references!}
  \typeout{******************************************}
  \typeout{}
 }

\end{document}